\documentclass[pra,showpacs,preprint,a4paper]{revtex4}
\usepackage{epsfig}
\usepackage{graphicx}
\usepackage{amsfonts}
\usepackage{color}
\usepackage{multirow}

\begin{document}

\title{Entropy, fidelity, and double orthogonality for resonance states in 
two-electron quantum dots}

\author{Federico M. Pont}
\email{pont@famaf.unc.edu.ar}
\affiliation{Facultad de Matem\'atica, Astronom\'{\i}a y F\'{\i}sica,
Universidad Nacional de C\'ordoba, and IFEG-CONICET, Ciudad Universitaria,
X5016LAE C\'ordoba, Argentina}
\author{Omar Osenda}
\email{osenda@famaf.unc.edu.ar}
\affiliation{Facultad de Matem\'atica, Astronom\'{\i}a y F\'{\i}sica,
Universidad Nacional de C\'ordoba, and IFEG-CONICET, Ciudad Universitaria,
X5016LAE C\'ordoba, Argentina}
\author{Julio H. Toloza}
\email{jtoloza@exa.unne.edu.ar}
\affiliation{Universidad Nacional del Nordeste, Avenida Libertad 5400, W3404AAS
Corrientes, Argentina}
\affiliation{IMIT--CONICET}

\author{Pablo Serra}
\email{serra@famaf.unc.edu.ar}
\affiliation{Facultad de Matem\'atica, Astronom\'{\i}a y F\'{\i}sica,
Universidad Nacional de C\'ordoba, and IFEG-CONICET, Ciudad Universitaria,
X5016LAE C\'ordoba, Argentina}

\begin{abstract}
{  Resonance states of a two-electron quantum dot are studied 
using a variational expansion with both real basis-set functions and 
complex scaling methods. The 
two-electron entanglement (linear entropy) is calculated as a function
of the electron repulsion at both sides of the critical value, where the
ground (bound) state becomes a resonance (unbound) state. The linear 
entropy and fidelity and double orthogonality functions are compared as methods
for the determination 
of the real part of the energy of a resonance. The complex linear
entropy of a resonance state is introduced using complex scaling formalism.}
\end{abstract}
\date{\today}

\maketitle
\section{Introduction}

In the past few years the application of quantum information concepts to some
long standing problems has led to a deeper understanding of those problems
\cite{osterloh2002} and,
as a consequence,
to the formulation of new methods to solve (or calculate) them. For example,
the simulability of many body problems is determined by the amount of
entanglement shared between the spins of the system \cite{schuch2008}.

There is a number of quantities that can be calculated in order to analyze the
information carried by a given state including the entanglement of formation
\cite{wootters1998},
the fidelity \cite{tano} and several kinds of entropies, entanglement witnesses
and so on.
Of course it depends on the problem which quantity is more adequate, or
accessible, to be calculated.  

In the case of atomic or few body systems with continuous degrees of freedom a
rather natural quantity is the von Neumann entropy, it has been used to study a
number 
of problems: Helium like atom \cite{osenda2007}, \cite{osenda2008}, 
generation of
entanglement via scattering \cite{schmuser2006}, the dynamical entanglement of
small molecules \cite{liu2008}, and   entanglement in  Hooke's
atom \cite{coe2008}. 

In quantum dots, most  quantum information studies focus in the amount of
entanglement carried by its eigenstates \cite{dot-states,abdullah2009}, or in
the
controllability of the
system \cite{control}. Both approaches are driven by the possible use of a
quantum dot as
the physical realization of a qubit\cite{loss1998}. The controllability of the
system is
usually investigated (or performed) between the states with the lowest lying
eigenenergies \cite{imamoglu1999}.

Besides the possible use of quantum dots as quantum information
devices there are proposals to use them as photodetectors. The proposal is
based on the use of resonance states because of its properties,
in particular its
large scattering section compared to the scattering sections of bounded states
\cite{sajeev2008}.

{ The resonance states are slowly decaying scattering states
characterized 
by a large but finite lifetime. Resonances are also signaled by sharp,
Lorentzian-type peaks in the scattering matrix. In many cases of interest where
complex scaling (analytic dilatation) techniques can be applied, resonances
energies show up as isolated complex eigenvalues of the rotated Hamiltonian
\cite{reinhardt1996}.} Under this
transformation the bound states remain exactly preserved and the resonance states are
exposed as  ${\cal L}^2$ functions of the rotated Hamiltonian. Resonance states
 can be observed in two-electron quantum dots \cite{bylicki2005,sajeev2008}
 and two-electron atoms \cite{dubau1998}. 

Recently, a work by Ferr\'on, Osenda, and Serra \cite{ferron2009} studied
the behavior of the
von Neumann entropy associated with ${\mathcal L}^2$
approximations
 of resonance states of two electron quantum dots. In particular that work
was focused in the resonance state that arises when the ground state {  loses} its
stability, {\em i.e.} the quantum dot does not have two electron bounded states
any more. { Varying} the parameters of the quantum dot allows the energy to cross the
ionization threshold that separates the region where the two electron ground
state is stable from the region where the quantum dot { loses} one electron.

In reference \cite{ferron2009} it was found that the von Neumann entropy
provide a way to obtain the real part of the energy of the resonance, in other
words, the von Neumann entropy provides a stabilization method. The numerical
approximation used in \cite{ferron2009} allowed to obtain only a reduced
number of energy levels (in a region
where the spectrum is continuous) their method provided the real part of the
energy of the resonance only for a discrete set of parameters and this set could
not be chosen {\em a priori}. Notwithstanding this, Ferr\'on {\em et al.},
conjectured that there
is a well defined function $S(E_r)$, the von Neumann entropy of the resonance
state, which has a well defined value for every value of the real part of the
energy of the resonance, $E_r$.

In this work we will show,  if $\lambda$ is the external
parameter that
drives the quantum dot through the ionization threshold, that the 
entropy $S(E_r(\lambda))$ is a smooth function of $\lambda$.  Also it is showed
that
the resonance state entropy calculated by Ferr\'on {\em et al.} is  correct near
the ionization
threshold.

We have studied other 
quantities, besides the entropy, that are good witnesses of resonance
presence.
One of them is the Fidelity, which has been widely used
\cite{zanardi2006,zanardi2007,zanardi2006,zanardi2009} in the detection of
non-analytical behavior in the spectrum of quantum systems. The analysis of the
Fidelity
provides a method to  obtain the real part of the resonance energy from 
variational eigenstates. We introduce the {\em Double
Orthogonality} function (DO) that 
measures  changes in  quantum
states and detects the resonance region. The DO compares the
 extended continuum states and the state near the resonance, also providing the
real part of the resonance energy.

The paper is organized as follows. In Section \ref{sec-two} we present the
model and briefly explain the technical details to obtain approximate
eigenvalues,
eigenfunctions, and the density of states for the problem. In Section
\ref{sec-three} the fidelity is used  to obtain the
real
part of the resonance energy and the Double Orthogonality is introduced as an 
alternative method. In Section~\ref{sec-four} the linear entropy and the
expectation value of the Coulombian repulsion are studied using complex scaling
methods. Finally, in Section~\ref{sec-conclu} we discuss our results and
present our conclusions.

\section{The Model and basic results}
\label{sec-two}

There are many models of quantum dots, with different symmetries and
interactions. In this work we consider a model with spherical symmetry, with
two
electrons interacting via the Coulomb repulsion. The main results should not be
affected by the particular
potential choice as it is already known that  the near  threshold
behavior and other critical
quantities (such as the critical exponents of the energy and other
observables)
are mostly determined by the range of the involved potentials 
\cite{pont_serra_jpa08}.  Therefore
to model the dot potential we use a short-range potential suitable to apply
the complex
scaling method. After this considerations we propose the following
Hamiltonian $H$ for the system

\begin{equation}
\label{hamiltoniano}
H = -\frac{\hbar^2}{2m} \nabla_{{\mathbf r}_1}^2  
-\frac{\hbar^2}{2m} \nabla_{{\mathbf r}_2}^2  + V(r_1)+V(r_2)+ 
\frac{e^2}{\left|{\mathbf r}_2-{\mathbf r}_1\right|} ,
\end{equation}
where $V(r)=-(V_0/r_0^2)\, \exp{(-r/r_0)}$, ${\mathbf r}_i$ the
position operator of electron $i=1,2$;   $r_0$ and $V_0$
determine  the range and depth of the dot potential.
After re-scaling with $r_0$, in atomic units the Hamiltonian of Eq.
(\ref{hamiltoniano}) can be written as
\begin{equation}
\label{hamil}
H = -\frac{1}{2} \nabla_{{\mathbf r}_1}^2  
-\frac{1}{2} \nabla_{{\mathbf r}_2}^2 -V_0 e^{-r_1}-V_0
e^{-r_2} + 
\frac{\lambda}{\left|{\mathbf r}_2-{\mathbf r}_1\right|} ,
\end{equation}
where $\lambda=r_0$.

We choose the exponential binding potential to take advantage of its analytical
properties. In particular this potential is well behaved and
the energy of the resonance states can be calculated using complex scaling
methods. So, besides its simplicity, the exponential potential allows us to
obtain independently the energy of the resonance state and a check for our
results. The threshold energy, $\varepsilon$, of Hamiltonian Eq.~(\ref{hamil}),
that is, the one body ground state energy can be calculated
exactly~\cite{galindo} and is given by \begin{equation}
J_{2\sqrt{2\varepsilon}}\left(\sqrt{2V_0}\right)=0,
\end{equation}
where $J_{\nu}(x)$ is the Bessel function.

The discrete spectrum {\bf and} the resonance states of the model given by 
Eq.  (\ref{hamil})  can be obtained approximately 
using ${\cal L}^2$
variational functions \cite{bylicki2005}, \cite{kruppa1999}. So, if
$\left|\psi_j(1,2)\right\rangle$ are the exact eigenfunctions of the
Hamiltonian, we look for variational approximations 

\begin{equation}\label{variational-functions}
\left|\psi_j(1,2)\right\rangle \,  \simeq\, 
\left|\Psi_j(1,2)\right\rangle \, =\, 
\sum_{i=1}^M c^{(j)}_{i} \left| \Phi_i 
\right\rangle \, ,\;\; c^{(j)}_{i} = (\mathbf{c}^{(j)})_i 
\;\;;\;\;j=1,\cdots,M \,.
\end{equation}

\noindent where the $\left| \Phi_i \right\rangle$ must be chosen adequately and $M$ is the
 basis set size. 

Since we are interested in the behavior of the system near the 
ground-state ionization
threshold, we choose as basis set  s-wave singlets given by

\begin{equation}\label{basis}
\left| \Phi_i\right\rangle \equiv \left| n_1,n_2;l\right\rangle = 
\left( \phi_{n_1}({r}_1) \, \phi_{n_2}({r}_2) \right)_s 
\mathcal{Y}_{0,0}^l (\Omega_1,\Omega_2) \, \chi_{s} \, ,
\end{equation}
where $n_2\leq n_1$, $l\leq n_2$, $\chi_{s}$ is the singlet spinor, 
 and the $\mathcal{Y}_{0,0}^l
(\Omega_1,\Omega_2) $ are given by
\begin{equation}\label{angular-2e}
\mathcal{Y}_{0,0}^l (\Omega_1,\Omega_2)\,=\, \frac{(-1)^l}{\sqrt{2l+1}} \,
\sum_{m=-l}^{l} (-1)^m Y_{l\,m}(\Omega_1) Y_{l\, -m}(\Omega_2) \, ,
\end{equation}
{\em i.e.} they are eigenfunctions of the total angular momentum with zero
eigenvalue and
the $Y_{l\, m}$ are the spherical harmonics. {  Note also that $\mathcal{Y}_{0,0}^l$
 is a real function since it is symmetric in the particle index.}  The  radial term
$(\phi_{n_1}({r}_1) \phi_{n_2}({r}_2))_s$
has the appropriate symmetry for a singlet state,

\begin{equation}\label{radial-sym}
(\phi_{n_1}({r}_1) \phi_{n_2}({r}_2))_s \,=\, \frac{\phi_{n_1}(r_1)
\phi_{n_2}(r_2)+ \phi_{n_1}(r_2) \phi_{n_2}(r_1)}{
\left[ 2\,(1+\langle n_1 | n_2
\rangle^2 ) \right]^{1/2}}
\end{equation}
\noindent where
\begin{equation}\label{int-prod}
\left\langle n_1|n_2 \right\rangle = \int_0^{\infty} r^2 \phi_{n_1}(r)
\phi_{n_2}(r) \, dr
\, \, ,
\end{equation}
\noindent and the $\phi$'s are chosen to satisfy $\langle n_1|n_1\rangle
= 1$. The numerical results  are obtained by taking
the Slater type forms for the orbitals
\begin{equation}\label{slater-type}
\phi^{(\alpha)}_{n}({r}) = \left[ \frac{\alpha^{2n+3}}{(2n+2)!}\right]^{1/2} r^n
e^{-\alpha r/2} .
\end{equation}
{  \noindent where $\alpha$ is a non-linear parameter of the basis.} It is clear that
 in terms of the functions defined in Eq. (\ref{basis}) the variational
 eigenfunctions reads as
\begin{equation}\label{variational-eigen}
\left|\Psi^{(\alpha)}_i(1,2)\right\rangle = \sum_{n_1 n_2 l} c^{(i),(\alpha)}_{n_1 n_2 l}
\left|
n_1, n_2;l;\alpha\right\rangle \, ,
\end{equation}
\noindent where $n_1\geq n_2\geq l \ge 0$, then the basis set size is given by
\begin{equation}
M = \sum_{n_1=0}^N \sum_{n_2=0}^{n_1} \sum_{l=0}^{n_2}  1 \,=\,
\frac{1}{6} (N+1) (N+2) (N+3)\; ,
\end{equation}
so we refer to the basis set size using both $N$ and $M$.  In Eq.
(\ref{variational-eigen}) we
 added $\alpha$ as a basis index to indicate that in general the
variational eigenfunction is $\alpha$-dependent.
The matrix elements of the kinetic energy, the Coulombic repulsion between the
electrons and other mathematical details involving the functions
$ \left| n_1, n_2 ;l;\alpha\right\rangle$ are given in
references~\cite{osenda2008b},
\cite{pablo-variational-approach}.  We only show here for completeness
 the matrix elements of the exponential potential in the basis of
Eq.~(\ref{slater-type}),

\begin{equation}\label{mat-exp}
\left\langle n\left|\, e^{-r} \,\right| n'\right\rangle
= \int^{\infty}_{0}
\phi_n(r)\phi_{n'}(r)\, e^{-r} \,r^2\,\textrm{d}r  = 
        \left(\frac{\alpha}{1+\alpha}\right)^{n+n'+3}\frac{(2+n+n')!}{\sqrt{
(2n+2)!\, (2n'+2)!}}.
\end{equation}

\begin{figure}[ht]
\begin{center}
\psfig{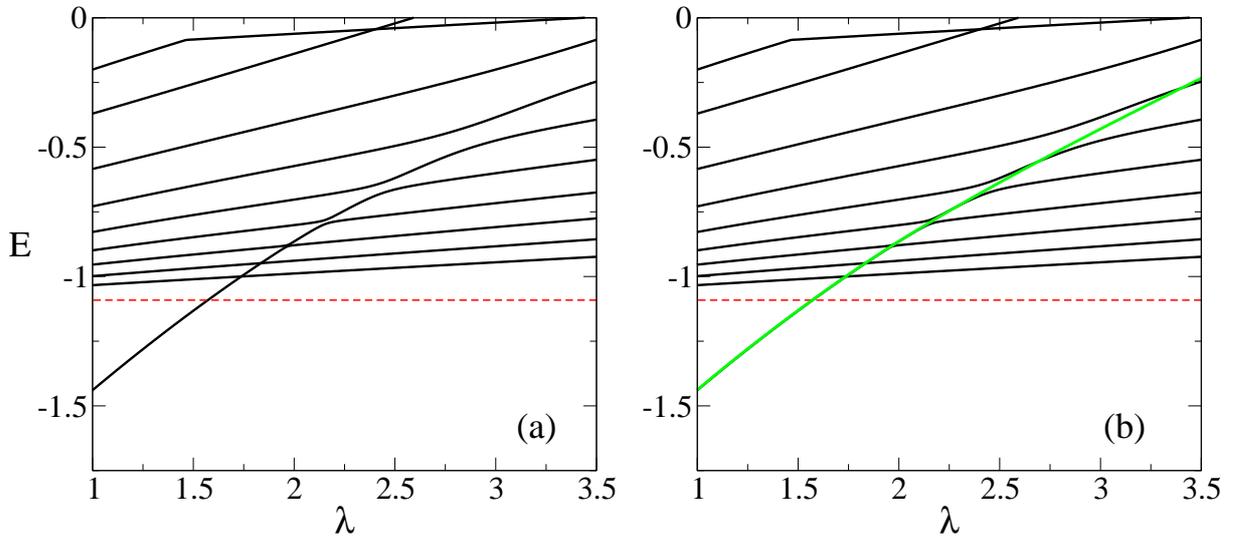}
\end{center}
\caption{\label{avoidedcross}(color on-line) (a) the figure shows the behavior
of the
variational eigenvalues $E_j^{(\alpha)}(\lambda)$ (black lines) for $N=14$ and non-linear parameter $\alpha=2$. The red dashed line
corresponds to the threshold energy $\varepsilon\simeq -1.091$.
Note that the avoided crossings between the variational eigenvalues
are fairly visible. (b) The figure shows the same variational eigenvalues that
(a) (black lines) and the energy calculated using the complex scaling method
(green line) for a parameter $\theta=\pi/10$. }
\end{figure}

Resonance states have isolated complex eigenvalues, $E_{res}=E_r -i \Gamma/2,\;
 \Gamma > 0$, whose eigenfunctions are not square-integrable.
These states are considered as quasi-bound states  of
energy $E_r$ and inverse life time  $\Gamma$. For the Hamiltonian Eq.
(\ref{hamil}), the resonance energies belong to the interval 
$(\varepsilon,0)$ \cite{reinhardt1996}.

The resonance states can be analyzed using the
spectrum obtained with a basis of ${\cal L}^2$ functions (see \cite{ferron2009} and
References therein). 
The levels above the threshold have several avoided crossings 
that ``surround'' the real part of the
energy of the resonance state.
The presence of a resonance
can be made evident looking at the eigenvalues
obtained numerically. Figure~\ref{avoidedcross} shows  a typical
spectrum obtained from the variational method. This
figure shows  the behavior of the variational eigenvalues
$E_j^{(\alpha)}$ as functions of the parameter $\lambda$. The results shown were
obtained using $N=14$ and $\alpha=2.0$. The value of $\alpha$ was chosen in
order to obtain the best approximation for the energy of the ground state in
the region of $\lambda$ where it exists. The Figure shows
clearly that for
$\lambda<\lambda_{th}\simeq1.54$ there is only one bounded state. Above the
threshold the variational approximation provides a finite number of solutions
with energy below zero. Above the threshold there is not a
clear cut criteria to choose the value of the variational parameter. However,
it is possible to calculate $E_r(\lambda)$  calculating $E_j^{(\alpha)}$ for many
different values of the variational parameter (see Kar and Ho \cite{kar2004}).

\begin{figure}[ht]
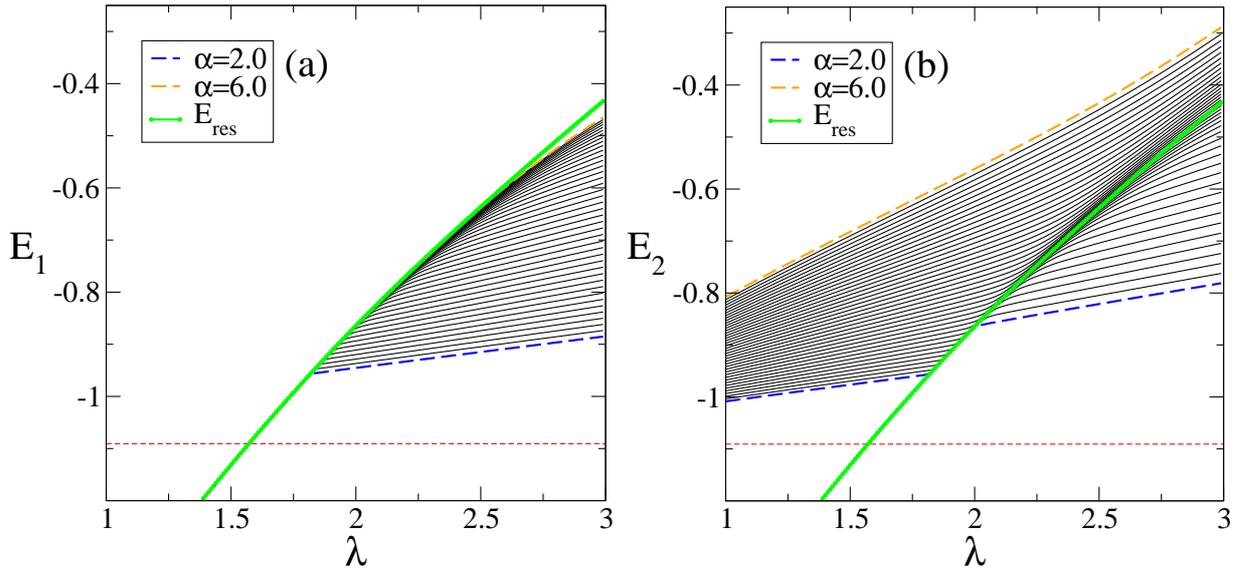

\begin{center}
\psfig{figure=fig2a_pont.eps,width=8cm}
\psfig{figure=fig2b_pont.eps,width=8cm}
\end{center}
\caption{\label{densidad-0-1}(color on-line) (a) The first variational
state energy {\em vs} $\lambda$, for different values of the variational
parameter $\alpha$. From bottom to top $\alpha$ increases its value from
$\alpha=2$ (dashed blue line) to $\alpha=6$ (dashed orange  line). The real part
of the resonance eigenvalue obtained using complex
scaling ( $\theta=\pi/10$) is also shown (green line). (b) Same as (a) but for the
second state energy.}
\end{figure}

Figure~\ref{densidad-0-1} (a) and (b) shows the numerical results for the first and second
eigenvalues respectively, for different values of the variational parameter
$\alpha$. 
The figure also shows the behavior of the ground state (below the threshold) and
the
real part of the energy of the resonance calculated using complex scaling
(above the threshold), this curve is used as a reference. The behavior of the
smaller variational eigenvalue $E_1^{(\alpha)}(\lambda)$ is rather clear, below the
threshold $E_1^{(\alpha)}(\lambda)$ is rather insensitive to the actual value of
$\alpha$, the differences between $E_1^{(\alpha=2)}(\lambda)$ and $E_1^{(\alpha=6)}(\lambda)$ are smaller
than the width of the lines shown in the figure. Above the threshold the
behavior changes, the curve for a given value of $\alpha$ has two well defined
regions, in each region $E_1$ is basically a straight line. The two straight
lines in each region has a different slope and  the change in the slope is
located around $E_r(\lambda)$.

In the case of $E_2^{(\alpha)}(\lambda)$ there are three regions, in each
one of them the
curve for a given value of $\alpha$ is basically a straight line and the slope
is different in each region. A feature that appears rather clearly is that, for
fixed $\lambda$ , the
density of levels for energy unit is not uniform,
despite that the
curves $E_j^{(\alpha_i)}(\lambda)$ are drawn for forty equally spaced
$\alpha_i$'s between $\alpha=2.0$ and $\alpha=6.0$. This
fact has been observed previously \cite{mandelshtam1993} and  the density of
states can be
written in terms of two contributions, a localized one and an extended one. The
localized density of states is attributed to the presence of the resonance
state, conversely the extended density of states is attributed to the continuum
of states between $(\varepsilon, 0)$.

\begin{figure}[ht]
\begin{center}
  \includegraphics[width=8cm]{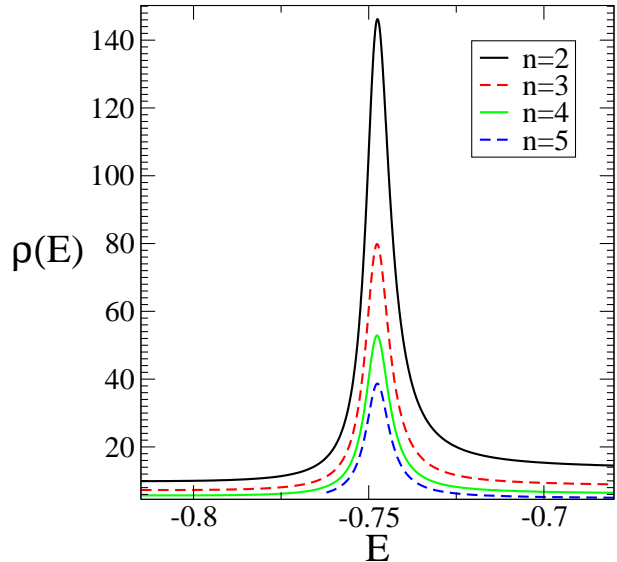}
\caption{\label{densidad} (color on-line) The density of states
$\rho(E)$ for  $\lambda=2.25$ and basis set size $N=14$. The results were
obtained using Eq. (\ref{densidad_sin_suma}) and correspond to, from top to
bottom, the second (black line), third (dashed red line), fourth (green line)
 and fifth (dashed blue line) levels.}
\end{center}
\end{figure}

The localized density of states $\rho(E)$ can be expressed as \cite{kar2004,mandelshtam1993}
\begin{equation}\label{densidad_sin_suma}
\rho(E)  = \left|\frac{\partial
E(\alpha)}{\partial \alpha}\right|^{-1} . 
\end{equation}
Since we are dealing with a variational approximation, we calculate
\begin{equation}\label{densidad_cal}
\rho(E_j^{(\alpha_i)}(\lambda))  = \left| 
\frac{E_j^{(\alpha_{i+1})}(\lambda) -
E_j^{(\alpha_{i-1})}(\lambda)}{\alpha_{i+1} - \alpha_{i-1}}\right|^{-1} .
\end{equation}
Figure~\ref{densidad} shows the typical behavior of $\rho_j(E)\equiv
\rho(E_j^{(\alpha_i)}(\lambda))$ for several
eigenenergies and $\lambda=2.25$. The real and imaginary parts of the
resonance's
energy, $E_r(\lambda)$ and $\Gamma$ respectively, can be obtained from
$\rho(E)$, see for example \cite{kar2004} and references there in. This method
 provides and independent way to obtain $E_{res}$, besides the method of
 complex scaling.

The values of $E_r(\lambda)$ and $\Gamma(\lambda)$ are obtained performing
a nonlinear fitting of $\rho(E)${, with a Lorentzian function,}

\begin{equation}
\rho(E)=\rho_0 + \frac{A}{\pi}\frac{\Gamma/2}{\left[(E-E_r)^2
+(\Gamma/2)^2\right]}.
\end{equation}

One of the drawbacks of this method results
evident: for each $\lambda$ there are several $\rho_j(E)$ (in fact one for each
variational level), and since each $\rho_j(E)$  provides a value for
$E^j_r(\lambda)$ and $\Gamma^j(\lambda)$ one has to choose which one is the
best.
Kar and Ho \cite{kar2004} solve this problem fitting all the $\rho_j(E)$ and
keeping as the best values for $E_r(\lambda)$ and $\Gamma(\lambda)$ the fitting
parameters with the smaller $\chi^2$ value. At least for their data the best
fitting (the smaller $\chi^2$) usually corresponds to the larger $n$. 
This fact has a clear interpretation, if the numerical method approximates
$E_r(\lambda)$ with $E^{(\alpha)}_n(\lambda)$ a large $n$ means that the numerical
method is able to provide a large number of approximate levels, and so the
continuum of states between $(\varepsilon,0)$ is ``better'' approximated. 

It is worth to remark that the results obtained from the complex scaling
method and from the density of states are in excellent
agreement, see Table~\ref{tabla1}.

\section{fidelity and double orthogonality functions}
\label{sec-three}

Since the work of Zanardi {et al.} \cite{zanardi2006,zanardi2007} there has
been a growing interest in the fidelity approach as a mean to study quantum
phase transitions \cite{zanardi2006}, the information-theoretic differential
geometry on quantum phase transitions (QPT's) \cite{zanardi2007} or
the disordered quantum $XY$ model \cite{zanardi2009}. In all these cases the
fidelity is used to detect the change of behavior of the states of a quantum
system. For example, if $\lambda$ is the external parameter that drives a system
through a QPT, the fidelity is the superposition ${\mathcal F} =
\langle\Psi(\lambda-\delta \lambda), \Psi(\lambda+\delta \lambda)\rangle $,
where $\Psi$ is the ground state of the system. It has been shown that ${\mathcal
F}$ is a good detector of critical behavior in ordered \cite{zanardi2006} and
disordered  systems\cite{zanardi2009}. 

In the following we will show that the
energy levels calculated using the variational approximation show critical
behavior near the energy of  the resonance, moreover the curve $E_r(\lambda)$
can be obtained from the fidelity.

\begin{figure}[ht]
\begin{center}
\psfig{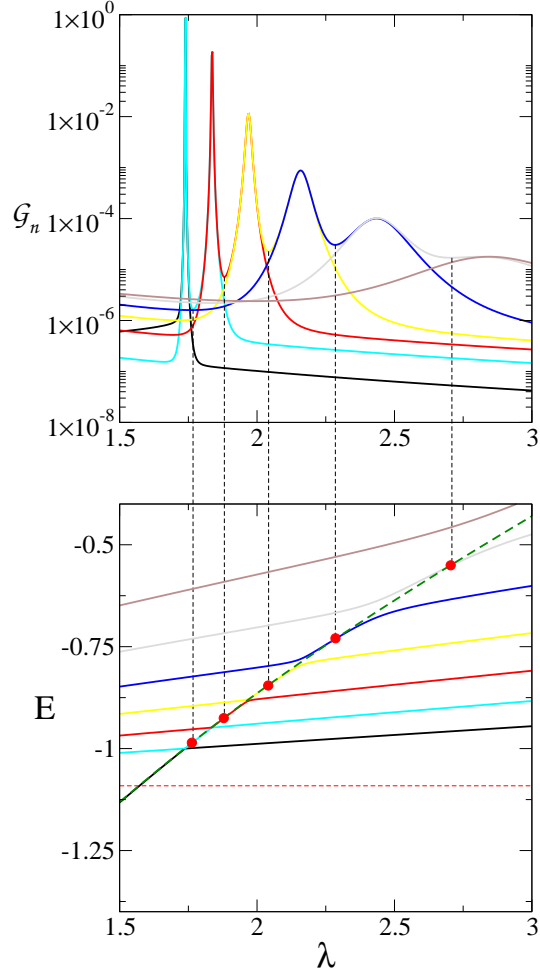}
\end{center}
\caption{\label{fidel1}(color on-line) The upper panel shows the behavior of
 ${\mathcal G}_n$, for $n=1, \ldots, 7$. Each function ${\mathcal G}_n$ has
two peaks, except for $n=1$. Since one of the peaks of ${\mathcal G}_n$
coincides with one of the peaks of ${\mathcal G}_{n+1}$, only one
peak for each level $n$ is apparent.  From left to right the visible line at
each peak correspond to $n=2, \ldots, 7$ (cyan, red, yellow, blue, grey, 
brown lines, respectively). The $n=1$ line has no visible peak (black). 
The lower panel shows the variational eigenlevels $n=1, \ldots, 7$ (with the 
same color convention used in the upper panel), and $E_{r}(\lambda)$ (green 
dashed line). The black vertical dashed lines connecting both panels show the
 value of $\lambda$  where each ${\mathcal G}_n$ has its minimum, $\lambda_n^f$.
 The red dots in the lower panel correspond to $E_n(\lambda^f_n)$.}
\end{figure}

Figure \ref{fidel1} shows the behavior of the function ${\mathcal G}_n= 1-F_n$,
where \\ $F_n
= |\langle\Psi_n(\lambda), \Psi_n(\lambda+\delta \lambda)\rangle|^2$ , and
$\Psi_n$ is the $n^{th}$ eigenstate obtained with the variational approach.

The behavior of ${\mathcal G}$ is quite simple. The value of ${\mathcal G}$ is
 very small,  except near the avoided
crossings where the value of ${\mathcal G}$ increases rather steeply (at least
for small $n$). This
is so because near the avoided crossing the superposition
$|\langle\Psi_n(\lambda), \Psi_n(\lambda+\delta \lambda)\rangle|^2\rightarrow
0$. Actually, $|\langle\Psi_n(\lambda),
\Psi_n(\lambda+\delta \lambda)\rangle|^2\rightarrow 0$ near points such that
$E^{(\alpha)}_n(\lambda)$ has non analytical behavior. Is for this reason that the
fidelity is a good detector of quantum phase transitions
\cite{zanardi2006,zanardi2007,zanardi2009}. In a first order QPT
the energy of the ground state is non analytical, and in a second order QPT the
gap in the avoided crossing between the ground state and the first excited
state goes to zero in the thermodynamic limit.

The previous argument supports why ${\mathcal G_1}$ has only one peak, while
all the others functions  ${\mathcal G_n}$ have two peaks, the number of peaks
is the number of avoided crossings of each level. However, since the resonance
state lies somewhere between the avoided crossings it is natural to ask what
feature of the fidelity signals the presence of the resonance. For a given
level $n$ the value of the energy is fixed, so we must pick a distinctive
feature of ${\mathcal G}_n$ that is present for some $\lambda^f_n$, such that
$ E_r(\lambda^f_n) \simeq E_n(\lambda^f_n)$ (from here we will use $E_n$ or
$E_n^{(\alpha)}$ interchangeably). It results to be, that
$\lambda^f_n$ is the value of $\lambda$ such that  ${\mathcal G}_n$ attains its
minimum between its two peaks. Figure~\ref{fidel1} shows the points
$E_n(\lambda^f_n)$. In Table ~\ref{tabla1} we tabulate the real part
of the energy calculated using DO, complex
scaling, fidelity and Density of States, for the five values of $\lambda^f_n$
shown in
Figure~\ref{fidel1}. The numerical values obtained using the fidelity and
Density of States methods are identical up to five figures. The Relative Error
between the energies 
obtained  is less than 0.25\%

\begin{table}[floatfix]
\caption{\label{tabla1} Resonance Energy obtained by four different methods. 
Basis size is N=14}   
\centering                          
\begin{tabular}{c c c c r r r r r}     
\hline\hline \\[-2.0ex]   
\multirow{2}{*}{\makebox[2.5cm][c]{$\lambda_{DO}^n$}} & &
\multirow{2}{*}{\makebox[2.5cm][c]{$DO$}}  &
\multirow{2}{*}{\makebox[2.5cm][c]{Complex}} &
\multicolumn{5}{c}{Fidelity and Density of States} \\
\cline{5-9}
       &          &          & \makebox[2.5cm][c]{ Scaling}& $n=2$    & $n=3$
     & $n=4$ &   $n=5$ &   $n=6$\\ [0.5ex] 
\hline
1.755 \scriptsize{(n=2)}  & E        & -0.99075  & -0.99098 & -0.99011   &   \----    & \----    & \----        & \----      \\
    & $\alpha$ & 2.0       &  2.0     &  1.560     &   \----    & \----   
&\----        & \----     \\
1.8625 \scriptsize{(n=3)} & E        &-0.93427   & -0.93452 & -0.93434   & -0.93383   & -0.93303 & \----        & \----         \\
 & $\alpha$ & 2.0       &  2.0     &   2.448    &  1.787     & 1.414   
&\----        & \----     \\
2.02  \scriptsize{(n=4)}  & E        &-0.85498   & -0.85556 & -0.85581   & -0.85564   & -0.85531 &	-0.85486     & \----         \\
       & $\alpha$ & 2.0       &  2.0     &   3.339    &  2.435     & 1.906   
&1.519        & \----     \\
2.255  \scriptsize{(n=5)}  & E        & -0.74329  & -0.74538 & -0.74518   & -0.74527   & -0.74519 &	-0.74521 & -0.74514 \\
       & $\alpha$ & 2.0       &  2.0     &   4.262    &   3.098    & 2.414   
&1.936    &  1.574   \\
2.61  \scriptsize{(n=6)}   & E        &-0.58276 & -0.59077 & -0.58825&-0.58910  &-0.58942  &-0.58965  & -0.58979\\
        & $\alpha$ & 2.0       &  2.0     &    5.248   &  3.799     & 3.799   
&2.373    &  1.936   \\  [1ex]         
\hline
\end{tabular}
\end{table}

The idea of detecting the resonance
state energy with functions depending on the inner product could be taken a
step further. To this end we consider the functions
\begin{equation}
\label{dort}
DO_n(\lambda) = |\langle \Psi_n(\lambda_{L}), \Psi_n(\lambda)   \rangle|^2 +
|\langle \Psi_n(\lambda_{R}), \Psi_n(\lambda)\rangle|^2, \mbox{for} \,
\lambda_{L} < \lambda < \lambda_{R}
\end{equation}
where $\lambda_{L}$ and $\lambda_{R}$ are two given coupling values.
It is clear from the definition that \mbox{$0 \leq DO_n(\lambda) \leq 2$}. 
If there are
not resonances between  $\lambda_{L}$ and $\lambda_{R}$, the wave function 
is roughly independent of $\lambda$ so $DO_n(\lambda)  \simeq 2$. 
However, the 
scenario is different when a resonance is present between  
$\lambda_{L}$ and $\lambda_{R}$. 
In this case, the  avoided crossings for a given state
$\Psi_n$  are located approximately at
$\lambda^{av}_{L}$
and $\lambda^{av}_{R}$, where $L$($R$) stands for the leftmost (rightmost)
 avoided crossing.  Requesting that
$\lambda_{L}<\lambda^{av}_{L}<\lambda^{av}_{R}<\lambda_{R}$, 
it follows that $\langle \Psi_n(\lambda_L)|
 \Psi_n(\lambda_R) \rangle \,\simeq\,0$.
 With this prescription, the $DO_n$ functions are rather independent of the actual
values chosen for $\lambda_{L}$ and
$\lambda_{R}$.
For a
given $n$,  $DO_n(\lambda)$ measures how much the state $\Psi_n(\lambda)$
differs from the extended states $\Psi_n(\lambda_{R})$ and
$\Psi_n(\lambda_{L})$. 

\begin{figure}[floatfix]
\begin{center}
\psfig{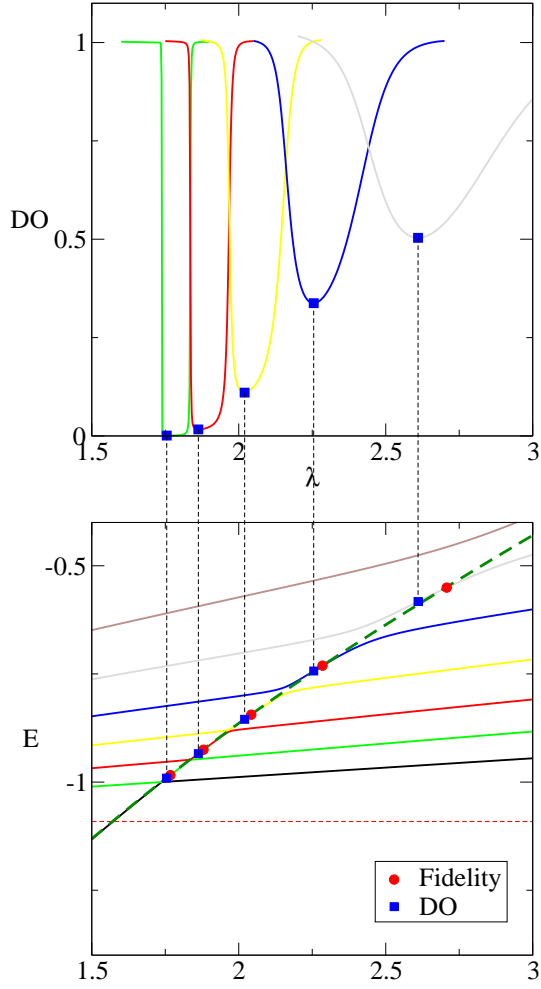}
\end{center}
\caption{\label{dortfig}(color on-line) The lower panel shows the
variational energy levels $E_{n}(\lambda)$, from bottom to top, for $n=1,\ldots,
 7$ (the black, green, red, yellow, blue, grey and brown continuous lines,
 respectively); $E_{r}(\lambda)$ (the dashed dark green line); 
$E_n(\lambda_{DO}^n)$ (blue squares) and $E_n(\lambda_{n}^f)$ (red dots). 
The upper panel shows the behavior of $DO_n$ {\em vs} $\lambda$ for 
$n=2,\ldots, 7$. The color convention for the $DO_n$ is the same used in the
 lower panel. The black dot-dashed vertical lines show the location of the
 points $\lambda_{DO}^n$. }
\end{figure}

We look for the states with minimum $DO_n$, in the same fashion as we did
with the fidelity, we can obtain values $E_r(\lambda^n_{DO}) \simeq
E_n(\lambda^n_{DO})$, where $\lambda^n_{DO}$ is defined by 
$DO_n(\lambda^n_{DO}) = \min_{\lambda} DO_n(\lambda)$. Figure~\ref{dortfig}
 shows the behavior of $DO_n$ obtained for
the same parameters that the ones used in Figure~\ref{fidel1}, and we compare
the values of  $E_n(\lambda^n_{DO})$ with energy values
obtained using complex scaling methods in Table~\ref{tabla1}. The curves in
Figure~\ref{dortfig} show that outside $(\lambda_{L},\lambda_{R})$ the states
$\Psi_n$ change very little when $\lambda$ changes and $DO_n \simeq 1$. Inside
the resonance region, $(\lambda_{L}^{av},\lambda_{R}^{av})$, the functions
$DO_{n}$ change
abruptly. The width in $\lambda$ in which a given $DO_n$ changes abruptly
apparently depends on the width of the resonance, $\Gamma$, but so far we have
not been
able to relate both quantities. 

From Table~\ref{tabla1} and Figure~\ref{dortfig} it is
rather clear that despite that the fidelity and the $DO_n$ provide approximate
values for $E_r(\lambda)$ for different sets of $\lambda$'s, both sets belongs
to the ``same'' curve, {\em i.e.} the same curve considering the numerical
inaccuracies. Both methods would give the same results when
$|\lambda^{av}_{R}-\lambda^{av}_{L} | \rightarrow 0$, but for $N$ finite the
fidelity measures how fast the state changes when
$\lambda\rightarrow\lambda+\Delta\lambda$ and the $DO$ measures how much a
state differs from the extended states located at both sides of the resonance
state.

\section{The entropy}
\label{sec-four}

If $\hat{\rho}^{red}$ is the reduced density operator for one electron
\cite{ferron2009}, then the
von
Neumann entropy ${\mathcal S}$ is given by
\begin{equation}\label{von-neumann-entropy}
{\mathcal S} = -\mathrm{tr}(\hat{\rho}^{\mathrm{red}}
\log_2{\hat{\rho}^{\mathrm{red}}}) ,
\end{equation}
and the linear entropy $S_{\mathrm{lin}}$ is given by \cite{abdullah2009}
\begin{equation}\label{linear-entropy}
{\mathcal S}_{\mathrm{lin}} = 1-\mathrm{tr}\left[(\hat{\rho}^{\mathrm{red}})^2
\right],
\end{equation}
where the reduced density operator is
\begin{equation}\label{rho-red-def}
\hat{\rho}^{\mathrm{red}}(\mathbf{r}_1, \mathbf{r}^{\prime}_1) =
\mathrm{tr}_2 \left| \Psi \right\rangle \left\langle \Psi \right| \, ,
\end{equation}
here the trace is taken over one electron, and  $\left|\Psi \right\rangle$ is
the total two-electron wave function. Both entropies,
Eqs. (\ref{von-neumann-entropy}) and (\ref{linear-entropy}), can be used to analyze
how much entanglement has a given state. One can choose between one entropy or
the other out of convenience. In this paper we will use the linear entropy. For
a discussion about the similarities between the two entropies see
Reference~\cite{abdullah2009} and references therein.

As the two electron wave function is not
available we instead use the variational approximation Eq. 
(\ref{variational-eigen}).
As has been noted in previous works (see \cite{osenda2007} and References
therein), when the total wave function factorizes in  spatial and spinorial
components it is possible to single out both contributions, then the analysis of
the behavior of the entropy is reduced to the analysis of the behavior of
the spatial part $S$, since the spinorial contribution is constant. In this
case, if $\varphi (\mathbf{r}_1, \mathbf{r}_2)$ is the two electron
wave function and $\rho^{red}(\mathbf{r}_1, \mathbf{r}^{\prime}_1)$ is given by

\begin{equation}
\rho^{red}(\mathbf{r}_1, \mathbf{r}^{\prime}_1) = \int
\varphi^{\star}(\mathbf{r}_1, \mathbf{r}_2) \varphi(\mathbf{r}^{\prime}_1,
\mathbf{r}_2) \; d\mathbf{r}_2 ,
\end{equation}
then the linear entropy $S_{\mathrm{lin}}$ can be calculated as
\begin{equation}
S_{\mathrm{lin}}= 1 -\sum_i \lambda_i^2 ,
\end{equation}
where the $\lambda_i$ are the eigenvalues of $\rho^{red}$ and are given by
\begin{equation}
\int \rho^{red}(\mathbf{r}_1, \mathbf{r}^{\prime}_1) \phi_i(\mathbf{r}^{\prime}_1)
\; d\mathbf{r}^{\prime}_1 = \lambda_i \phi_i(\mathbf{r}_1) \, .
\end{equation}

\begin{figure}[floatfix]
\begin{center}
\psfig{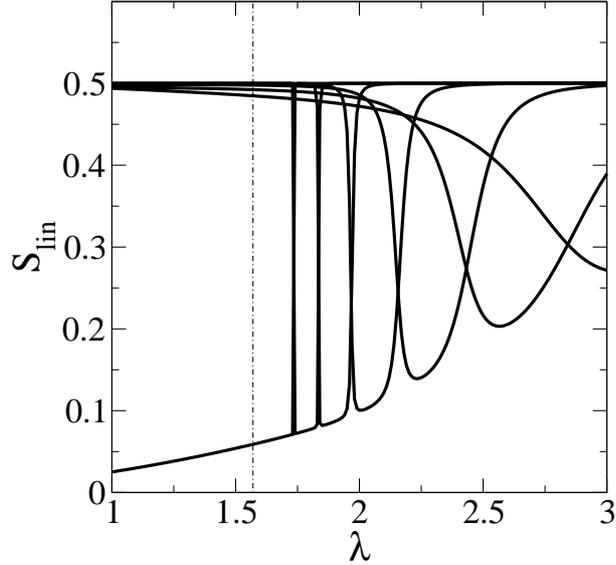}
\end{center}
\caption{\label{entro-lin} The figure shows the behavior of
$S_{\mathrm{lin}}(\Psi_j^{(\alpha)})$, where  the $\Psi_j^{(\alpha)}$ are the variational 
eigenstates corresponding to the first seven energy levels shown in Figure
\ref{avoidedcross} for $N=14$ and $\alpha=2{.}0$. All the curves $S_{\mathrm{lin}}(\Psi_j^{(\alpha)})$ , except for
the corresponding
to $S_{\mathrm{lin}} (\Psi_1^{(\alpha)})$, have a single minimum located at
$\lambda_j^S$, {\em i.e.} $S_{\mathrm{lin}}(\Psi_j^{(\alpha)}(\lambda_j^S)) =
\min_{\lambda} S_{\mathrm{lin}}(\Psi_j^{(\alpha)}(\lambda)) $ . If $i<j$ 
then $\lambda_i^S < \lambda_i^S$. }
\end{figure}

Figure \ref{entro-lin} shows the behavior of the linear
entropy for several variational levels. The meaning of each curve has been
extensively discussed in Reference \cite{ferron2009}. We include a brief
discussion here for completeness.

When the two-electron quantum dot loses an electron the state of
the system can be
described as one electron
bounded to the dot potential, and one unbounded electron {\em at infinity}, as
a consequence the
spatial wave function can be written as a symmetrized product 
of one electron wave functions so $S_{\mathrm{lin}}=S_c=1/2$.  Therefore if only
bound and continuum states are considered the entropy has a discontinuity
when $\lambda$ crosses the threshold value $\lambda_{th}$. The  picture
changes  significantly when resonance states are considered.
The resonance state keeps its two electrons
``bounded'' before the ionization  for a finite time given by the inverse of
the imaginary part of the energy. Of course the  life time of a bounded
state is infinite while the life time of a resonance state is finite. In
reference \cite{ferron2009} is suggested
 that it is possible to construct a smooth function
$S(E_r(\lambda))$ that ``interpolates'' between the minima of the functions
$S(\Psi_j)$ shown in Figure~\ref{entro-lin}. This assumption was justified by
similar arguments that the  used in the present work, {\em i.e.} if we
call
$\lambda_n^S$ to the value of $\lambda$ where $S(\Psi_n)$ gets its minimum then
$E_n(\lambda_n^S)$ follows approximately the curve $E_r(\lambda)$. As
Ferr\'on {\em et al.} \cite{ferron2009} used only one variational parameter
$\alpha$, it
seemed natural to pick the minimum value of $S(\Psi_n)$ as the feature that
signaled the presence of the resonance state. 

Until now we have exploited the fact that $E_r(\lambda)$, at a given $\lambda
$ can be approximated by variational eigenvalues corresponding to different
values of the variational parameter, say $E_r(\lambda) \simeq
E_{n}^{(\alpha)}(\lambda) \simeq E_{n^{\prime}}^{(\alpha^{\prime})}(\lambda)$ 
(the
superscript $\alpha$ is made evident to remark that the eigenvalues correspond
to different variational parameters $\alpha$ and $\alpha^{\prime}$ ). There is
no problem in approximating an exact eigenvalue with different variational
eigenvalues. But, from the point of view of the entropy, there is a problem 
since, in general, $S(\Psi_n^{(\alpha)}(\lambda)) $ is not close to
$S(\Psi_n^{(\alpha^{\prime})}(\lambda))$. Moreover, as has
 been stressed in Reference~\cite{schuch2008}, a given numerical method could be
useful to
 accurately calculate the spectrum of a quantum system, but hopelessly
 inaccurate to calculate the entanglement. In
few body systems there is evidence that there is a strong correlation between
the entanglement and the Coulombian repulsion between the components of the
system
\cite{osenda2007,coe2008,abdullah2009,osenda2008b}. Because of this correlation
we will carefully investigate the behavior of the Coulombian repulsion between
the electrons in our model. 

For the Hamiltonian Eq. (\ref{hamil}), and $\psi \in {\cal L}^2$ an eigenvector of $H$ with
eigenvalue $E$, the Hellman-Feynman theorem gives that
\begin{equation}\label{h-f}
\frac{\partial E}{\partial \lambda} = \left\langle \psi \right|
\frac{1}{r_{12}}\left| \psi\right\rangle.
\end{equation}
We use both sides of Eq. (\ref{h-f}) to analyze how the variational
approximation works with expectation values of observables that are not the
Hamiltonian. The $r.h.s$ of Eq. (\ref{h-f}) is well defined if we
use ${\cal L}^2$ functions as the approximate variational eigenfunctions.

\begin{figure}[floatfix]
\begin{center}
\psfig{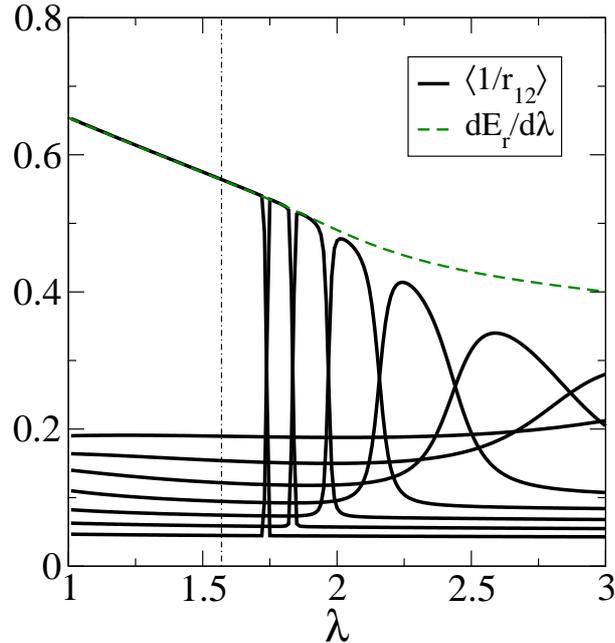}
\end{center}
\caption{\label{hell-fey} The figure shows the expectation values of the Coulombian
 repulsion for the variational states $\Psi^{(\alpha)}_n$, $n=1,\ldots,8$ with
$N=14$ and $\alpha=2.0$. Also is showed the curve $\frac{\partial E_r}{\partial \lambda}$ 
obtained from the complex Energy of the complex scaling method.}
\end{figure}

To evaluate the  $l.h.s$  of Eq. (\ref{h-f}) we take advantage that we
have found, independently, the real part of the
resonance eigenvalue, $E_r(\lambda)$,  using complex
scaling methods. Figure~\ref{hell-fey} shows the behavior of
$\frac{dE_r}{d\lambda}$ and the Coulombian repulsion between the two
electrons, $\langle \frac{1}{r_{12}} \rangle_n$, where
$\langle \ldots\rangle_n$ stands for the expectation value calculated with
$\Psi^{(\alpha)}_n$. The behavior of $\langle \frac{1}{r_{12}}\rangle_n$ is
quite simple to analyze, where the linear entropy of $\Psi^{(\alpha)}_n$ 
has a valley the
expectation value $\langle \frac{1}{r_{12}} \rangle_n$ has a { peak}.
Where the expectation value $\langle \frac{1}{r_{12}}
\rangle_n$ has its maximum the corresponding linear entropy has its
minimum. The inverse behavior showed by the entropy and the Coulombian repulsion
has been observed previously \cite{abdullah2009,osenda2008b}. 
 
For a given variational parameter $\alpha$, and for small $n$, $\langle
\frac{1}{r_{12}} \rangle_n$ has its maximum very close to the curve
$\frac{dE_r}{d\lambda}$. Besides, the shape of both curves near the maximum of
$\langle\frac{1}{r_{12}} \rangle_n$ is very similar, in this sense our
variational approach gives a good approximation not only for $E_r(\lambda)$ but
for its derivative too.

\begin{figure}[floatfix]
\begin{center}
\psfig{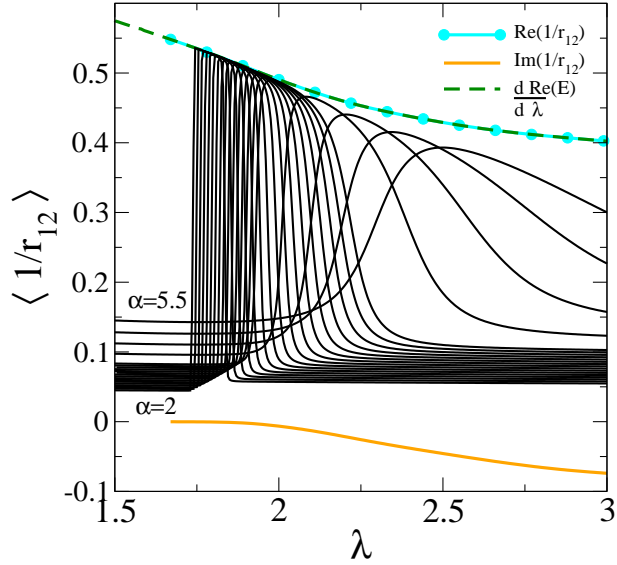}
\end{center}
\caption{\label{r12-a-n}(color-online) The figure shows the expectation value 
$\langle 1/r_{12} \rangle_2^{(\alpha)}$ vs. $\lambda$ for a basis size $N=14$ and $\alpha=2,\,\ldots\,,3.5$
in $0.1$ steps and for $\alpha=4,\,\ldots\,,5.5$ in $0.5$ steps (solid black 
lines). The real (cyan dotted) and imaginary (orange line) 
parts of  $\langle 1/r_{12} \rangle_\theta$  ($\theta=\pi/10$), 
and the derivative of the
real part of the complex-scaled energy are also shown. }
\end{figure}

For larger values of $n$ the maximum of $\langle
\frac{1}{r_{12}} \rangle_n $ gets apart from the curve of
$\frac{dE_r}{d\lambda}$, and the shape of the curves near this maximum is quite
different. We proceed as before and changing $\alpha$ we obtain a  good
approximation for $\frac{dE_r}{d\lambda}$ up to a certain value
$\lambda_{rep}$. For any $\lambda$ smaller than $\lambda_{rep}$, there
is a pair
$n,\alpha$ such that $\langle
\frac{1}{r_{12}} \rangle_{n,\alpha}$ is locally close to
$\frac{dE_r}{d\lambda}$ and the slope of both curves is (up to numerical
errors) the same, see figure~\ref{r12-a-n}.

Apparently there is no way to push further the variational method, at least
keeping the same basis set, in order to obtain a
better approximation than the depicted in Figure~\ref{r12-a-n}. The
difficulty seems to be more deep than just a limitation of the variational
method used until this point. We can clarify this subject using the properties
of the complex scaling method. Let us call $\phi^{\theta}$ the eigenvector
such that
\begin{equation}\label{complex-eigen}
H(\theta) \phi^{\theta} = E_{res} \phi^{\theta} ,
\end{equation}
where $H(\theta)$ is the Hamiltonian obtained from the complex scaling
transformation \cite{moisereport}, and $\theta$ is the angle of ``rotation''.
The eigenvector $\phi^{\theta}$ depends on $\theta$, but for $\theta$ large
enough the eigenvalue $E_{res}$ does not depend on $\theta$. As pointed by
Moiseyev \cite{moisereport}, the real part of the expectation value
of a complex scaled observable is the physical measurable quantity, while the
imaginary part gives the uncertainty of measuring the real part. Moreover,
the
physical measurable quantity must be $\theta$ independent as is, for example,
the eigenvalue $E_{res}$.

The eigenvector $\phi^{\theta}$ can be normalized using that
\begin{equation}\label{complex-norm}
\langle (\phi^{\theta})^{\star} |\phi^{\theta} \rangle =1.
\end{equation}
Since $\phi^{\theta}$ is normalized, we get that
\begin{equation}\label{hell-fey-exp}
\frac{\partial E_{res}}{\partial \lambda} = \left\langle
(\phi^{\theta})^{\star} \right| \frac{e^{-i\theta} }{r_{12}} \left|\phi^{\theta}
\right\rangle = \left\langle\frac{1}{r_{12}} \right\rangle_{\theta},
\end{equation}
in this last equation we have used that, under the complex scaling
transformation, 
\begin{equation}
\frac{1}{r_{12}}\rightarrow \frac{e^{-i\theta}}{r_{12}},
\end{equation}
and defined the quantity $ \left\langle\frac{1}{r_{12}}
\right\rangle_{\theta}$. This generalized Hellman-Feynman theorem is also 
valid for Gamow states \cite{hfg}.

Figure~\ref{r12-a-n} shows the behavior of the expectation value in the
$ \left\langle\frac{1}{r_{12}}
\right\rangle_{\theta}$  as a function of
$\lambda$. It is clear that the real part of the expectation value $
\left\langle\frac{1}{r_{12}}
\right\rangle_{\theta}$
coincides with $\frac{\partial E_{r}}{\partial \lambda}$. More interestingly,
$\lambda_{rep}$ is where the imaginary part of $ \left\langle\frac{1}{r_{12}}
\right\rangle_{\theta}$ became noticeable. From this fact we conclude that it
is not possible to adequately approximate the Coulombian repulsion of a
resonance state, or its entropy, only with real ${\cal L}^2$ variational functions
despite its success when dealing with the resonance state spectrum.

We define the complex scaled density operator of the resonance state by
\begin{equation}\label{complex-rho}
\rho^{\theta} = \left|\phi^{\theta} \right\rangle \left\langle
(\phi^{\theta})^{\star} \right|,
\end{equation}
and the complex linear entropy
\begin{equation}\label{complex-entropy}
S^{\theta} = 1 - {\mathrm tr} (\rho^{\theta}_{red})^2 ,
\end{equation}
where
\begin{equation}\label{complex-reducida}
\rho^{\theta}_{red} = {\mathrm tr}_2 \rho^{\theta}, 
\end{equation}
and $\phi^{\theta}$ is the eigenvector of Eq. (\ref{complex-eigen}). { 
This definition is motivated by the fact that the density operator  should be 
the projector onto  the
space spanned by $\left|\phi^{\theta}\right\rangle$. As the normalization Eq.
 (\ref{complex-norm}) requires the {\em bra} to be conjugated, then $\rho^\theta$ is
the adequate projector to use.} 

Because
of the normalization, Eq. (\ref{complex-norm}), we have that ${\mathrm tr}
\rho^{\theta} = {\mathrm tr}\rho^{\theta}_{red}=1$ {\em despite that both
density operators have complex eigenvalues}. 

\begin{figure}[floatfix]
\begin{center}
\psfig{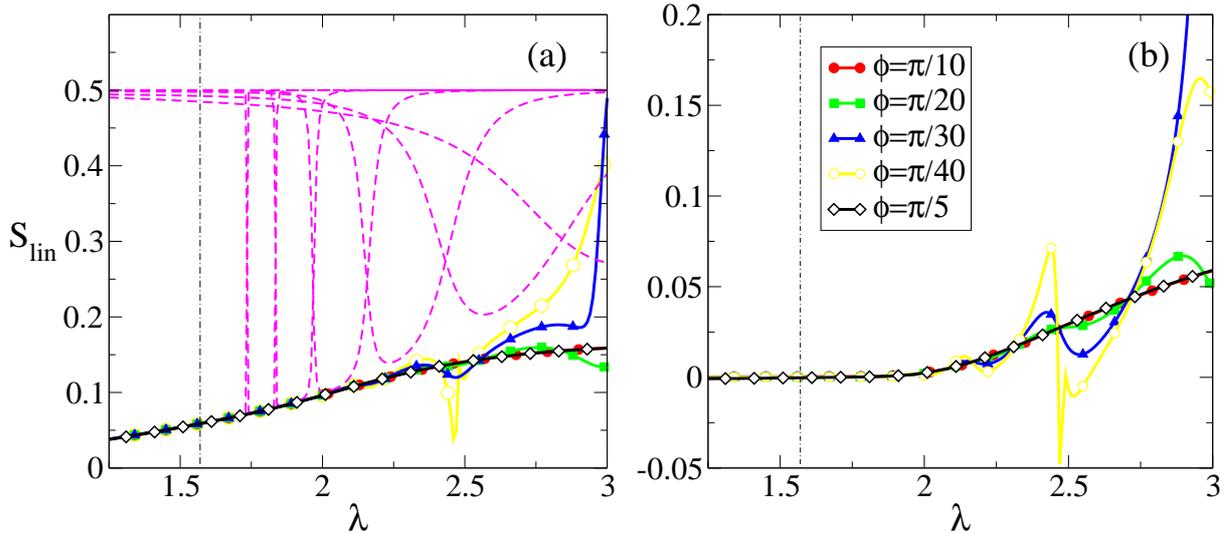}
\end{center}
\caption{\label{fig-complex-entropy}(color-online) Figure (a) shows the linear
entropy for the same values as in figure \ref{entro-lin} (magenta dashed lines).
 Also showed is the real part of the complex linear entropy for several values
of the complex rotation angle $\theta=\frac{\pi}{5},\, \frac{\pi}{10},\, \frac{\
pi}{20},\, \frac{\pi}{30},\, \frac{\pi}{40}$ (black empty diamonds, red dots, 
green squares, blue triangle, yellow empty dots). 
Figure (b) shows the imaginary part of the complex linear entropy for the same 
values as in (a). }
\end{figure}

Figure~\ref{fig-complex-entropy} shows that up to certain value of $\lambda$
the real part of $S^{\theta}$ follows closely a envelope containing the minima
of the functions $S(\Psi_n)$. However, for $\lambda$ large enough, $S^{\theta}$
gets apart from the functions $S(\Psi_n)$. It is worth to mention that for
$\theta$ large enough $S^{\theta}$ {\bf does not depend on} $\theta$. On the
other hand, far away from the threshold, the complex scaling requires 
larger values of $\theta$ to isolate the resonance state eigenenergy, but in
this regime the method becomes unstable. Because of the numerical evidence,
near the threshold the entropy calculated  by Ferr\'on {\em et al.} 
is basically correct, but for larger values of $\lambda$ the amount of 
entanglement of the
resonance state  should be characterized by $S^{\theta}$ and not by any of the
$S(\Psi_n)$. 

\section{summary and conclusions}
\label{sec-conclu}
We have presented numerical calculations about the behavior of the fidelity
and the double orthogonality functions $DO_n(\lambda)$. The numerical results
show that it is possible to obtain $E_r(\lambda)$ with great accuracy, for
selected values of $\lambda$, without employing any stabilization method.
These two methods to find  $E_r(\lambda)$ do not depend on particular
assumptions about the model or the variational method used to find approximate
eigenfunctions above the threshold, their success depends on the ability of the
approximate eigenstates to detect the non-analytical changes in the spectrum.

The fidelity has been extensively used to detect quantum phase
transitions in spin systems \cite{zanardi2006}, the behavior of
quasi-integrable systems \cite{weinstein2005}, thermal phase transitions
\cite{quan2009}, etc. To the best of our knowledge this work is the first
attempt to apply the concept of fidelity to resonance states and to the
characterization of spectral properties of a system with non-normalizable
eigenstates. Besides, it is remarkable that the fidelity and the double
orthogonality give the real part of the resonance eigenvalue using only real
variational functions. This energy as a function of $\lambda$ is obtained by
 moving the nonlinear parameter $\alpha$ but
{\em without} fitting as needed by standard
stabilization methods. Moreover, as shown by the tabulated values in
Table~\ref{tabla1} the fidelity provides $E_r(\lambda)$ as accurately as the
density of states method, with considerable less numerical effort.

We proposed a definition of the  resonance entropy
based on a complex scaled extension of the usual definition. The extension
implies that the reduced density operator is not hermitian and has complex
eigenvalues, resulting in a complex entropy.
The real and imaginary
parts of the complex entropy are $\theta$ independent, as should be expected for
the expectation value of an observable \cite{moisereport}. This
independence gives support to the interpretation of the real part of the entropy
as the amount of entanglement of the resonance state. 

Other kinds of resonances, as those  that arise from  perturbation of 
bound states embedded in 
the continuum, could be studied  applying the same quantum information methods 
used in this paper. Work is in progress in this direction.

\acknowledgments
We would like to acknowledge  SECYT-UNC,  CONICET and FONCyT 
for partial financial support of this project.


\begin{thebibliography}{20}


\bibitem{osterloh2002}A. Osterloh, L. Amico, G. Falci, and R. Fazio, Nature {\bf
416}, 608 (2002).

\bibitem{schuch2008} Norbert Schuch, Michael M. Wolf, Frank Verstraete, and
J. I. Cirac, Phys. Rev. Lett. 100, 030504 (2008).

\bibitem{wootters1998} W. K. Wootters, Phys. Rev. Lett {\bf 80}, 2245
(1998).

\bibitem{tano}Longyan Gong and Peiqing Tong, Phys. Rev. B {\bf 78}, 115114
(2008); Wing-Chi Yu, Ho-Man Kwok, Junpeng Cao, and Shi-Jian Gu,
Phys. Rev. E {\bf 80}, 021108 (2009); David Schwandt, Fabien Alet, and Sylvain
Capponi,
Phys. Rev. Lett. {\bf 103}, 170501 (2009).


\bibitem{osenda2007} O. Osenda and P. Serra, Phys. Rev. A {\bf 75}, 
042331 (2007).

\bibitem{osenda2008} O. Osenda and P. Serra, J. Phys. B: At. Mol. Opt. Phys.
{\bf 41}, 065502 (2008).


\bibitem{schmuser2006}F. Schm\"user and Dominik Janzing, Phys. Rev. A {\bf 73},
052313 (2006).

\bibitem{liu2008}Yan Liu, Yujun Zheng, Weiyi Ren, and Shiliang Ding, Phys. Rev.
A {\bf 78}, 032523 (2008).

\bibitem{coe2008}J. P. Coe, A. Sudbery, and I. D'Amico, Phys. Rev. B {\bf 77},
205122 (2008).

\bibitem{dot-states} Clive Emary, Phys. Rev. B {\bf 80}, 161309(R) (2009);
Robert Roloff and Walter P\"otz, Phys. Rev. B 76, 075333 (2007).

\bibitem{abdullah2009} S. Abdullah, J.P. Coe, and I. D'Amico, Phys. Rev. B {\bf
80}, 235302 (2009).

\bibitem{control}L. S\ae{}len, R. Nepstad, I. Degani, and J. P. Hansen, Phys.
Rev. Lett. {\bf 100}, 046805 (2008).

\bibitem{loss1998}Daniel Loss and David P. DiVincenzo, Phys. Rev. A {\bf 57},
120 (1998).

\bibitem{imamoglu1999} A. Imamo\={g}lu, D.D. Awschalom, G. Burkard, D.P.
DiVincenzo, D. Loss, M. Sherwin, and A. Small, Phys. Rev. Lett. {\bf 83}, 4204
(1999).


\bibitem{reinhardt1996}W. P. Reinhardt and Seungsuk Han, International Journal
of Quantum Chemistry {\bf 57}, 327 (1996).

\bibitem{sajeev2008}Y. Sajeev, and N. Moiseyev, Phys. Rev. B {\bf 78}, 075316
(2008).

\bibitem{bylicki2005}M. Bylicki, W. Jask\'olski, A. Stach\'ow, and J. Diaz,
Phys. Rev. B {\bf 72}, 075434 (2005).



\bibitem{dubau1998}J. Dubau, and I. A. Ivanov, J. Phys. B: At. Mol. Opt. Phys.
{\bf 31} 3335 (1998).


\bibitem{ferron2009}A. Ferr\'on, O. Osenda and P. Serra, Phys. Rev. A {\bf 79},
032509 (2009).

\bibitem{zanardi2006}P. Zanardi and N. Paunkovi\'c, Phys. Rev. E {\bf 74}, 031123
(2006).

\bibitem{zanardi2007}P. Zanardi, P. Giorda, and M. Cozzini, Phys. Rev. Lett.
{\bf 99}, 100603 (2007).

\bibitem{zanardi2009} S. Garnerone, N. T. Jacobson, S. Haas, and P. Zanardi,
Phys. Rev. Lett. {\bf 102}, 057205 (2009).

\bibitem{kruppa1999} A.T. Kruppa and K. Arai, Phys. Rev. A {\bf 59}, 3556
(1999).

\bibitem{pont_serra_jpa08} F. M. Pont and P. Serra, J. Phys A: Math. Theor. {\bf 41}, 275303 (2008).

\bibitem{galindo} A. Galindo and P. Pascual. {\em Quantum Mechanics I} ( Eudema, 1989).

\bibitem{osenda2008b}O. Osenda, P. Serra and S. Kais, International Journal of
Quantum Information {\bf 6}, 303 (2008).

\bibitem{pablo-variational-approach} P. Serra and S. Kais, Chem. Phys.
Lett. {\bf 372}, 205 (2003).

\bibitem{kar2004}Sabyasachi Kar and Y. K. Ho, 
J. Phys. B: At. Mol. Opt. Phys. {\bf 37}, 3177 (2004).



\bibitem{moisereport} N. Moiseyev, Phys. Rep. {\bf 302}, 211(1998). 

\bibitem{hfg} P. Ziesche, K. Kunze and B. Milek, J. Phys. A: Math. Gen. {\bf 20}, 2859(1987).

\bibitem{weinstein2005}Yaakov S. Weinstein and C. S. Hellberg, Phys. Rev. E
{\bf 71}, 016209 (2005).

\bibitem{quan2009}H. T. Quan and F. M. Cucchietti, Phys. Rev. E {\bf 79}, 031101
(2009).


\bibitem{mandelshtam1993} V. A. Mandelshtam, T. R. Ravuri and H. S. Taylor, Phys. Rev. Lett. {\bf 70},
1932 (1993).
\end{thebibliography}
\end{document}